# Magnetic Storm-substorm Relationship and Some Associated Issues

E. P. Savov

Solar-Terrestrial Influences Laboratory, Bulgarian Academy of Sciences
Acad. G. Bonchev St., Block 3, Sofia 1113, Bulgaria
E-mail: eugenesavov@mail.orbitel.bg

**Abstract**

The impinging solar wind and its magnetic field perturbed the Earth's magnetosphere and create magnetic storms and substorms. The Earth's magnetosphere expands (contracts) during periods of southward (northward) IMF. It is shown that these magnetospheric expansions and contractions account for poorly understood aspects of magnetic storm-substorm relationship, bifurcation of the magnetotail and the appearance of theta aurora. Quantitative theory and calculations in agreement with the suggested model of solar wind/IMF-magnetosphere coupling are presented. Pre-noon and post-noon dents on the magnetopause are expected to appear during a long period of strong northward IMF. The contracting and expanding magnetosphere accumulates solar wind plasma and its magnetic field. The accumulated energy is delivered into the polar regions and creates auroral activity intensification.

**Introduction**

The strength and duration of the southward interplanetary magnetic field (IMF) component are connected with the occurrence of geomagnetic storms and substorms. The IMF has to exceed a threshold value in order to induce or deepen a magnetic storm [1]. Magnetic substorms occur when the average southward IMF component $B_z < -3$ nT persists for one hour [2]. More than three hours duration of steady southward IMF ($B_z < -10$ nT) causes intense ($D_{st} < -100$ nT) magnetic storms [3]. The magnetic substorm consists of a growth phase, which is characterised with equatorward shifting of the boundaries of the auroral oval after southward IMF turning. Then the substorm expansion phase follows and poorly understood multiple surge forms are created on the poleward edge of the auroral oval as it contracts toward the quiet-time condition after the IMF turns northward [4]. Magnetic storms occur during longer periods of steady southward IMF and are characterised with a global decrease of the horizontal geomagnetic field component at middle and low latitudes. This is indicated by negative $D_{st}$ excursion, associated with a larger equatorward shift of the boundaries of the auroral oval during the main phase of storm than for the growth phase of substorm. The global geomagnetic field depression weakens and the auroral oval contracts after the IMF turns northward during the recovery phase of the magnetic storm [e.g. 3, 5]. The boundaries of the auroral field-aligned current system also shift equatorward (poleward) during periods of southward (northward) IMF [6].

The expansion phase of substorm occurs for any IMF orientation but it is most intense during southward IMF created main phase of magnetic storm [5]. There are poorly understood indications for near Earth location of the substorm expansion onset [7, 8] and the relationship between magnetic storms and substorms is controversial [9-11]. The mechanism of substorm expansion onset is debatable [11] and it is "quite difficult to differentiate between small magnetic storms and large substorms" [9].

Magnetic reconnection [12] is the basic paradigm of the solar wind/IMF-magnetosphere coupling. The three-dimensional (3D) magnetospheric circulation between the dayside magnetopause and the magnetotail, associated with reconnection, runs into a topological crisis [13]. This confusing situation is mended by smaller scale reconnection processes, called percolation reconnection [13, 14]. Small-scale processes are unlikely to maintain the self-consistency of the global magnetospheric convection. The threshold strength of prolonged southward IMF necessary to induce or deepen magnetic storm [1] implies a global character of solar wind/IMF-magnetosphere interaction, which is difficult to discuss in local reconnection terms. The global nature of this interaction is described with southward (northward) IMF driven expansion (contraction) of the magnetosphere accounting for the creation and dissipation of magnetic storms and substorms and their relationship [15]. The magnetic expansions and contractions will be discussed here in terms of 3D-spiral magnetic reconfiguration driven by the solar wind its magnetic field (i.e. the IMF).

It is believed that the substorm expansion phase ejects particles for the storm-time ring current whose intensification deepens the global low and middle latitude depression of the geomagnetic field horizontal component and so makes the $D_{st}$ index more negative. Hence substorms have to increase the magnitude of magnetic storm as measured by $-D_{st}$ and the beginning of the main phase $D_{st}$ decrease should occur after at least one substorm expansion onset. Anyway observations showed the opposite. Substorms are not a necessary condition for magnetic storm [9]. Substorms decrease the magnitude of a storm and the rate of its increase [16] and $D_{st}$ begins to decrease "well before" the first main phase substorm expansion onset [17].

The found expansions and contractions of the magnetosphere [15] are considered in 3D-spiral magnetic reconfiguration terms. It will be shown that the solar wind and the IMF driven 3D-spiral magnetic reconfiguration accounts for the controversial magnetic storm-substorm relationship, the longer growth phase of isolated substorm, the bifurcation of the magnetotail, the appearance of transpolar arc (i.e. theta aurora) and sunward convection. The role of the solar wind dynamic pressure in predicting the main phase $D_{st}$ excursion and the existence of a threshold value for the strength of the southward IMF that leads to initiation or deepening of magnetic storm will be also explained in the terms of the 3D-spiral magnetic reconfiguration. Appearance of pre-noon and post-noon dents on the magnetopause together with theta aurora is predicted. This work demonstrates that the 3D-spirally contracting and expanding magnetosphere accumulates energy from the solar wind and afterwards releases this energy to smaller scales within its structure.

## Magnetic expansions and contractions

Fig. 1 shows that the southward IMF enters directly into the northern polar region, exits from the southern polar region and so expands the magnetosphere in a way similar to the interaction between two attracting magnetic field sources. The solar wind is repelled at inner magnetospheric lines in the expanded magnetic configuration. So the impinging solar wind creates a larger magnetotail and thus the boundaries of the auroral oval shift equatorward during the expansion of the magnetosphere produced by southward IMF (Figs. 1, 2). The ionospheric roots of the inner lines of the magnetospheric configuration are equatorward from those of its outer lines. Hence the repelled from the inner lines particles will create auroral precipitation equatorward from that produced by particles which are repelled at outer lines. Thus the expansion (contraction) of magnetosphere accounts for the observed expansion (contraction) of the auroral oval during periods of southward (northward) IMF (Figs. 1-3).

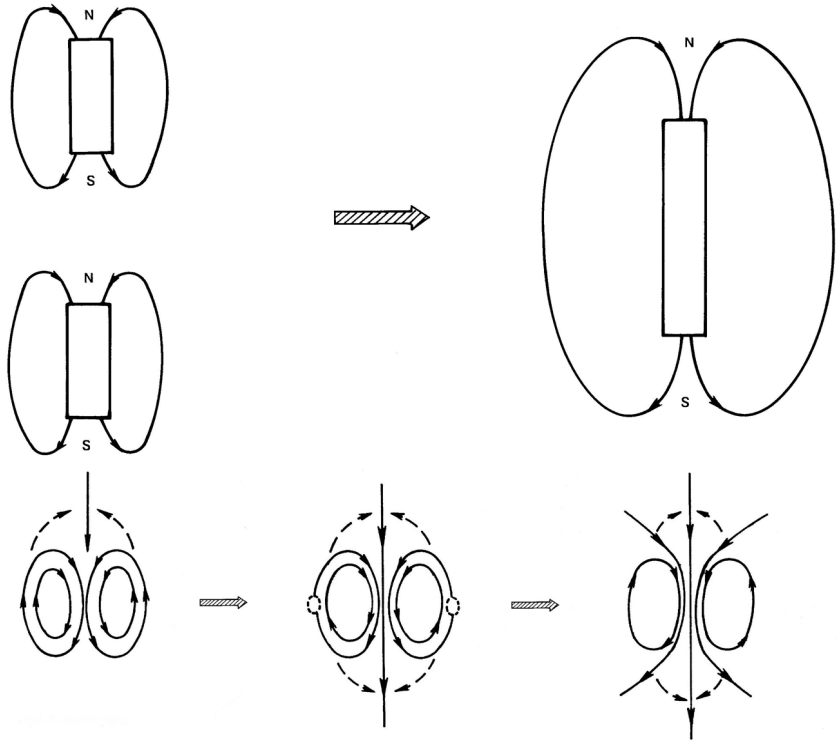

**Fig. 1.** Two sources of magnetic field face their opposite poles and attract. Then the external field, e.g., the southward IMF component ($B_z < 0$), enters directly into the magnetic configuration through its northern polar region and exits from the southern polar region. So the configuration expands as lines of magnetic force move 3D-spirally outward (the dashed arrows), oppositely to their inward 3D-spiral motion during the magnetic contraction shown in Fig. 3. The lines are carried downstream into the magnetotail by the impinging solar wind and become open (closed) for larger (smaller) magnetic expansion.

It can be seen from Fig. 1 that prolonged and steady (less prolonged and weaker) southward IMF will expand more (less) the magnetosphere. The greater expansion is indicated by a) appearance of low and middle latitude depression of the geomagnetic field, known as main phase of magnetic storm, b) decrease of the distance to the dayside magnetopause and c) creation of a thicker magnetotail and more equatorward shifted boundaries of the auroral oval. The smaller expansion creates the growth phase of substorm.

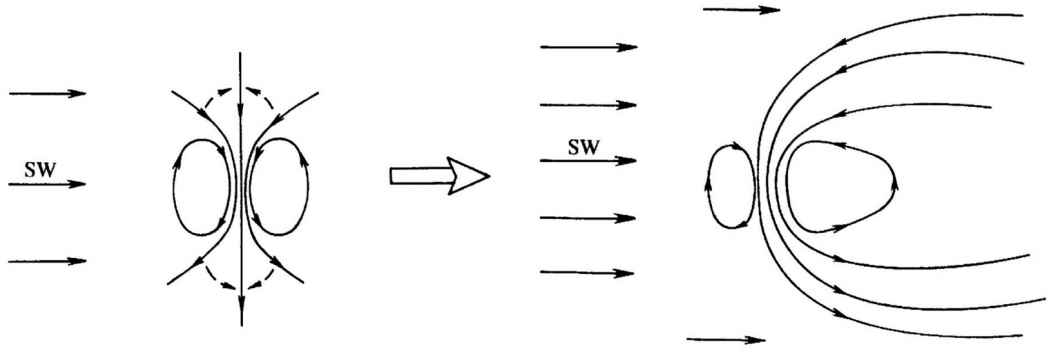

**Fig. 2.** The impinging solar wind (SW) expands the magnetosphere downstream. Thus it creates the downstream tilted lines of the magnetotail and the upstream tilt belt around them. Particle precipitation is delivered near the tilt reversal, which is at lower (higher) latitudes in the more (less) expanded part of magnetosphere. In this way downstream (night side) tilted polar cap surrounded by auroral oval are created. The plasma sheet is produced between the lobes of the magnetotail and the magnetospheric expansions and contractions energise its particles.

The boundaries of the auroral oval will shift equatorward (poleward) during expansion (contraction) of the magnetosphere for southward (northward) IMF (Figs. 1-3). So tailward propagating partial contractions of the magnetotail will create the puzzling near Earth location of the substorm expansion onset and the poorly understood appearance of multiple surge forms on the poleward edge of the auroral oval during the expansion phase of the substorm.

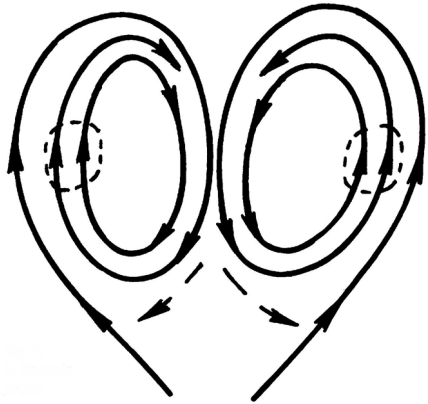

**Fig. 3.** Two sources of magnetic field face their same poles and repel. Then the external field, e.g., the northward IMF ($B_z > 0$), enters indirectly into the northern polar region of the magnetic configuration and winds up 3D spirally into it. The configuration contracts (the dashed arrows) and the intensity of its field increases at low and middle latitudes (the dashed curves). Thus the solar wind is repelled at outer magnetic lines and a greater distance to the dayside magnetopause, smaller magnetotail and smaller auroral oval are created. In the contracting (expanding) magnetic configuration the entering field winds up 3D-spirally inward into (unfolds 3D-spirally outward from) its source.

The spiral magnetic field **B** structure (Fig. 3) is likely to show zero flux through every closed surface at the finite, i.e. not arbitrary high, accuracy of magnetic interaction measurement. That is why the spiral magnetic structure allows *div***B** = 0 description. The topological crisis of the global magnetic circulation [13] resulting from this description is avoided in the proposed 3D-spiral magnetic expansions and contractions (Figs. 1-3). The topological crisis is created as the magnetic lines of force cross during their 3D circulation between the dayside magnetopause and the magnetotail [13]. The proposed 3D-spiral contractions and expansions of the magnetosphere (Figs. 1-3) are topological crisis free. The suggested 3D-spiral magnetic reconfiguration (Figs. 1-3) avoids the topological crisis, which is created from the mathematical properties of the zero divergence field whose lines are closed. The mathematical properties of the model should not be imposed on reality. The proposed qualitative model of the solar wind-magnetosphere interaction (Figs. 1-3) accounts for many confusing observations. It is also confirmed with quantitative assessments.

Figs. 1 and 2 show, in an agreement with observations [2, 3], that the growth phase of a substorm or the main phase of a storm are produced, respectively, from smaller or larger expansion of the magnetosphere, created during a shorter or longer period of weaker or stronger southward IMF. The inability to differentiate between a strong magnetic substorm and weak storm [9] is explained in this way. The substorm expansion phase is created from tailward propagating partial contraction of the magnetosphere. So in accordance with observations substorm expansions, usually called simply substorms or auroral substorms, will occur for different IMF orientations and will decrease the magnitude of magnetic storm. Thus the expansion phase of substorms, in agreement with the perplexing experimental findings [16], will weaken the negative $D_{st}$ excursion or will decrease the rate of its increase. The tailward spreading partial contraction of the magnetospheric configuration creates the well-known auroral features of the substorm expansion phase. It begins with brightening of the most equatorward auroral arc and afterwards a bulge appears on the poleward edge of the auroral oval due to delivery of auroral precipitation to higher latitudes in the contracting part of the magnetosphere. In this scenario the magnetospheric expansion, seen as beginning of magnetic storm main phase $D_{st}$ decrease, should occur in accordance with the confusing observations [17], i.e. well before the first tailward spreading partial contraction of the magnetosphere that creates the first main phase substorm expansion onset.

The IMF southward (northward) component expands (contracts) the magnetosphere (Figs. 1-3). Therefore the puzzling unusually long growth phase of an isolate substorm [18] was observed because it had taken more time for the southward IMF to expand the more contracted magnetosphere, created during the previous long period of steady northward IMF (Fig. 3).

Fluctuations in the persisting southward IMF and changes in the solar wind dynamic pressure are likely to produce more violent tailward spreading contractions, seen as more intense substorms, which occur during a greater expansion of the magnetosphere, i.e. during the main phase of the storm. This explains why the most intense substorms are observed during the main phase of magnetic storm.

The southward IMF enters into the northern polar region and expands the magnetosphere (Fig. 1). This directly accounts for the main phase $D_{st}$ decrease before ring intensification due to earthward ejected particles from the contracting parts of the magnetotail and poorly understood upward acceleration of ionospheric plasma. The lack of carries for the auroral field-aligned current system [19, 20] is accounted with transverse magnetic disturbances created by the upstream tilt belt of the main field around the downstream tilted lines of the magnetotail (Fig. 2). The insufficiency of carriers for the ring current [21] is directly explained with expansion of magnetosphere creating the negative storm-time $D_{st}$ excursion (Figs. 1, 2). The closely associated patterns of transverse to the main field magnetic disturbances, electric fields, plasma convection (i.e. $E \times B$ drift) and also field-aligned acceleration and precipitation in the high latitude ionosphere [e.g. 22, 23] are likely to be produced from lines of 3D-spiral magnetic reconfiguration. These lines will generate transverse and aligned magnetic disturbances and electric fields. Plasma follows the motion of the lines of magnetic reconfiguration (Figs. 1-3) and thus drifts tiltward at high latitudes and accelerates upward at low latitudes. The upward acceleration provides ionospheric particles for the ring current during magnetic expansion seen as main phase of storm.

**Quantitative assessments**

It takes on average 45 minutes for the auroral oval to expand after southward IMF turning and 8 hours to contract after the IMF becomes northward [24]. The typical velocity of the solar wind impinging on the Earth's magnetosphere is about 460 km/s. Then it takes 45 minutes for the IMF to be carried to about $195R_E$ downstream, where $R_E$ = 6,378 km is the radius of the Earth. In this way the southward IMF enters directly into the northern lobe of the magnetotail, exits from the southern lobe and thus expands the magnetosphere faster (Figs. 1, 2) than it contracts during the indirect northward IMF entry (Figs. 1-3). So the magnetosphere expands or contracts to extent that corresponds to the strength of the entering external field. Therefore in agreement with observations [1] there must be a threshold value for the strength of the entering southward IMF that will initiate or deepen the magnetospheric expansion, which is seen as beginning or deepening of magnetic storm. The strength of persisting negative (positive) IMF $B_z$ component defines the scale of magnetospheric expansion (contraction), which defines the average distance to the dayside magnetopause and the global low latitude depression of the geomagnetic field measured by the $D_{st}$ index (Figs. 1-3).

Let in accordance with observations [3] sustained IMF $B_z$ = −10 nT persisting for more than 3 hours creates magnetic storm having $D_{st}$ = -100 nT. The average distance to the dayside magnetopause is $R_m = 10R_E$ and the width of the magnetotail is about $2R_m = 20R_E$. Solar wind having average velocity of 460 km/s will carry during more than 3 hours the IMF to more than $780R_E$ downstream. In this way the southward IMF engulfs the northern lobe of the magnetotail and begins to pour into it and to expand the magnetosphere as shown in Figs. 1 and 2. Then the external magnetic flux entry is larger than $195R_E \times 20R_E \times 10$ nT, which is larger than the poleward transferred magnetospheric flux $|D_{st}| \times \pi(R_m)^2 = 100$ nT $\times 3.14 \times (10R_E)^2$ that creates the global low latitude depression of the geomagnetic field associated with magnetic storm (Figs. 1, 2). Therefore the southward IMF entry through the northern lobe of the magnetotail can drive the magnetospheric expansion indicated by the main phase negative $D_{st}$ development.

The length of the magnetotail depends on the volume of the southward IMF flux entry and hence is connected with the depth of the negative $D_{st}$ excursion (Figs. 1, 2). The greater the magnetotail the larger the IMF entry. This makes the magnetotail even greater until it cannot be maintained by the source of the geomagnetic field. Then the magnetotail collapses, i.e. undergoes contractions that deliver the energy accumulated from the solar wind to smaller scales. Therefore the magnetosphere reconfigures to enhance its interaction with the solar wind and thus to acquire energy from it and "to digest" it in its expanding and contracting structure.

Prolong and steady northward IMF contracts strongly magnetosphere and creates upstream (sunward) tilt in the noon sector of the polar cap (Fig. 3). Then lines around the upstream tilt, i.e. the magnetospheric lines from the pre-noon and post-noon sectors of the magnetopause, will be carried downstream with the solar wind. This will create pre-noon and post-noon dents on the magnetopause, magnetotail having four lobes, particle precipitation and transpolar arc (theta aurora) between the lobes. The puzzling theta aurora [e.g., 25] appears from considerable magnetospheric contraction that leads to bifurcation of the magnetotail (Figs. 3 and 2). The solar wind and the IMF driven 3D-magnetic reconfiguration creates regions of main field shear that separate the parts of the magnetosphere and are associated with particle precipitation and auroral displays.

The negative (positive) $D_{st}$ index is a quantitative measure of the magnetospheric expansion (contraction) due to the impinging solar wind and its IMF $B_z$ component (Figs. 1-3). Let $\rho$ is the density of the solar wind and $V$ is its velocity at the Earth's orbit. Then coupling function of the type $\rho V^2 B_s$ where $B_s = -B_z$ when $B_z < 0$ and $B_s = 0$ for $B_z > 0$ will correlate well with $D_{st}$. The found better $D_{st}$ prediction for $B_s V^2$ than for $B_s V$ [26] and the greater success of the coupling functions $p^{1/3} V B_s$ and $p^{1/2} V B_s$ [27-29] are in good agreement with the proposed coupling function, which also reveals the role of the solar wind dynamic pressure $p = \rho V^2$ (Figs. 1-3). The small negative background level of $D_{st}$ [5] indicates small persisting expansion of the magnetosphere due to the prevalence of the expanding factors – the solar wind dynamic pressure and the IMF $B_z < 0$ (Figs.1-3). It can be seen that the IMF $B_z > 0$ is the only factor that that contracts the magnetosphere as a whole (Figs. 1-3).

The IMF Bz < 0 expands the magnetosphere and the IMF Bz > 0 contracts it. Similar is the magnetospheric response to higher or lower solar wind dynamic pressure because of creation, respectively of larger or smaller magnetotail (Figs. 1-3). The magnetosphere expands and contracts as a whole and in parts and so undergoes self-similar transforms. This self-similarity allows the behaviour of each of its parts to be mapped to that of the others.

Let the external magnetic flux $\Phi_e$ that enters the northern polar region of the magnetic configuration is given by $\Phi_e = B_e \times S$, where $B_e$ is the entering magnetic flux density through a surface $S$, which is perpendicular to the entering field. Then the larger the external flux that enters into the magnetosphere the faster is its reconfiguration, i.e. the faster is the magnetospheric expansion or contraction (Figs. 1-3). Hence

$t_e \sim 1/S_e$, $t_c \sim 1/S_c$ and $t_e/t_c = S_c/S_e$,  (1)

where $t_e$ and $t_c$ are the times, respectively, for magnetic expansion and contraction, $S_e$ is the area which is perpendicular to the entering southward IMF that expands the magnetosphere and $S_c$ is the area perpendicular to the northward IMF entry which contracts it.

Let $R_m$ is the average distance to the dayside magnetopause and $d_m$ is the width of the low latitude magnetopause boundary layer through which the northward IMF flows into the northern polar region as shown in Figures 3 and 2. Then the area of the southward IMF entry is $S_e = \pi(R_m - d_m)^2$, assuming that it enters through a circle having radius $(R_m - d_m)$, which is in a plane over the northern polar region and the plane is perpendicular to the entering southward IMF (Figs. 1, 2). The northward IMF enters through area $S_c = (\pi(R_m)^2 - \pi(R_m - d_m)^2)$, which is perpendicular to the entering flux as it finds its way into the magnetopause boundary layer toward the northern polar region (Figs. 3, 2). The ratio between the both areas is $S_e/S_c = \pi(R_m - d_m)^2/(\pi(R_m)^2 - \pi(R_m - d_m)^2) = t_c/t_e$ - equal to the ratio between the times for the IMF driven contraction and expansion of the magnetic configuration (Eq. 1). Hence the created self-similar expansions and contractions of the magnetic configuration allow writing

$d = R[1 - (1/(M_{ec} + 1))^{1/2}]$,  (2)

where $R$ is the distance to the interface between the outer magnetosphere and the ambient space plasma in magnetic equatorial plane, $d$ is the width of the boundary layer created at the interface in the equatorial plane due to magnetic expansions and contractions, $M_{ec} = t_e/t_c$, called magnetic contraction rate, is the ratio between the expansion and contraction times $t_e$ and $t_c$ of the magnetic configuration. Equation (2) gives the width of the $d$-layer, which is created between the space plasma and the low latitude outer magnetosphere.

Figures 1 and 2 show that the increase (decrease) of solar wind dynamic pressure creates larger (smaller) magnetotail and so expands (contracts) the magnetosphere similarly to the southward IMF increase (decrease). The self-similar expansions and contractions of the magnetosphere (Figs. 1-3) suggest that Equation (2) holds for all external conditions and magnetic local time sectors. Therefore the increase of the magnetic contraction rate, e.g. during the expansion phase of substorm or during sudden solar wind dynamic pressure decrease, will make the space plasma-magnetosphere interface wider. Then intensification of precipitation created from earthward pulled plasma in the magnetic local time sector of the magnetic contraction will be observed. The self-similar response of the contracting and expanding as a whole and in parts magnetosphere allows Equation (2) to be applied to the magnetosphere as a whole and also to every of its magnetic local time sectors.

The mean time of the auroral oval expansion after southward IMF turning is 45 minutes and the oval contracts on average for 8 hours (480 minutes) after the IMF becomes northward [24]. Then from Equation (2) the average magnetospheric contraction rate is $M_{ec} = t_e/t_c = (45$ minutes$)/(480$ minutes$) = 0.094$ and the average thickness of the dayside low latitude magnetopause boundary layer is $d = 0.044R = 0.44R_E$ for $R = R_m = 10R_E$ equal to the average distance to the dayside magnetopause. The thickness of the dayside magnetopause boundary layer vary from 0 km to more than 1 $R_E$ [e.g. 30 and reference therein]. The discovered IMF $B_z$ component driven expansions and contractions of the magnetosphere [15] account for this variation (Eq. (2)) and the calculated average thickness of the dayside low latitude magnetopause boundary layer is consistent with it. The magnetic contraction rate is $M_{ec} \gg 1$ during fast magnetic contraction. Then the width of the $d$-layer is $d < R = 10R_E$. During fast magnetic expansion $M_{ec} \ll 1$ and the width of the $d$-layer is $d \approx 0$. The faster the magnetic contraction (expansion) the larger (the smaller) the width of the $d$-layer. Plasma from this layer is pushed inward into the contracting magnetic configuration. This leads to intensification and poleward expansion of the particle precipitation seen as brightening and poleward expansion of the auroral arc associated with the magnetic contraction, e.g. during the substorm expansion onset. The increase of the local contraction rate is likely to account for brightening of auroral arcs at all local times.

Equation (2) shows that the average width of the low latitude magnetopause boundary layer decreases with the decrease of the magnetic contraction rate $M_{ec}$ that is with the increase of the magnetospheric expansion. The decrease of the width of the magnetopause boundary layer is expected in the expanded configuration because in it the solar wind dynamic pressure is balanced at inner and hence stronger magnetic field lines. The field becomes stronger closer its source, i.e. the intensity the field increases inward into its configuration. Hence the magnetopause boundary layer will become thicker (thinner) in the IMF $B_z > 0$ (IMF $B_z < 0$) contracted (expanded) magnetospheric configuration because the solar wind particles create a thicker (less thicker) magnetopause boundary by penetrating deeper (less deeper) into the weaker (stronger) field. Variations of the thickness of the dayside magnetopause boundary layer can be produced by the inevitable IMF fluctuations, which will create multiscale contractions and expansions of magnetosphere (Figs. 1-3) and hence variations of the magnetic contraction rate $M_{ec}$ (Eq. (2)). The reported puzzling oscillations of the magnetopause above the level expected from the solar wind dynamic pressure variations [31] confirm this magnetospheric behaviour. Solar wind and IMF driven changes in the magnetic contraction rate will lead to changes in the auroral activity, which will intensify in the contracting local time sectors of the magnetosphere, e.g. during the expansion phase of substorms.

The expansion and contraction of the magnetosphere (Figs. 1-3) can be tested with the laboratory experiment described in Figure 4. Here the famous Birkeland terrella experiment that shows the formation of artificial auroral oval [e.g. 32] is upgraded with simulation of the IMF. This is done to demonstrate the expansion and contraction of the artificial auroral oval according the polarity of the applied external magnetic field, which simulates the IMF $B_z$ component. The IMF was discovered with spacecraft observations in the second half of the twentieth century. The crucial role of the IMF in the creation of magnetic substorms and storms [e.g. 2, 3] was not known by the time Birkeland performed his terrella experiment in end of nineteenth century.

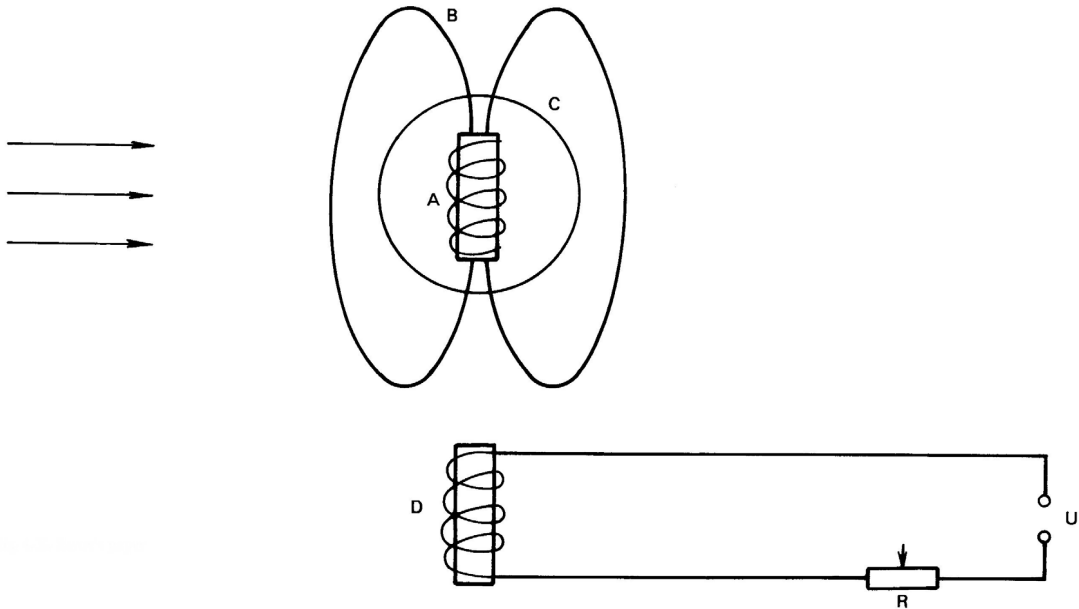

**Fig. 4.** Laboratory simulation of the solar wind/IMF driven expansions and contractions of the magnetosphere. Electrons shown by the three arrows impinge on the magnetic configuration (B), generated by the magnetic coil (A), which is inserted into the sphere (C) that is coated with a fluorescent paint. The magnetic poles of the field created from the coil (A) are put on one straight line with the magnetic poles of the field generated from the secondary coil (D), that is connected to a circuit containing the rheostat (R) and the direct current source (U). The impinging electrons have velocities parallel to the equatorial plane of the magnetic configuration (B). The field created from the magnetic coil (D) simulates the IMF.

Topological crisis is created from crossing of the magnetic field lines of force [13]. The topological crisis free 3D-spiral magnetic contractions and expansions account for magnetic field-plasma interactions in space and laboratory, and elucidate poorly understood issues of the storm-substorm relationship and solar wind-magnetosphere coupling (Figs. 1-4).

The 3D magnetic reconfiguration (Figs. 1-3) predicts that the obtained artificial auroral oval should expand (contract) when the magnetic configurations generated from the two coils (Fig. 4) meet their opposite (same) magnetic poles. The artificial auroral oval will expand (contract) when the two magnetic field sources that simulate the Earth's magnetosphere and the IMF attract (repel) as shown in (Figs. 1-4).

The observed huge vortices (about $6R_E$ across) at the magnetopause [33] are likely to originate as the collisionless space plasma follows the multiscale 3D-spiral magnetospheric contractions and expansions through which the Earth's magnetosphere acquires energy from its environment and delivers it to smaller scales. Plasma follows the lines of the 3D-spiral magnetic reconfiguration, which in the case of magnetospheric contraction (IMF $B_z > 0$) drives plasma inward to fill the plasmasphere. The latter will be depleted during magnetic expansion (IMF $B_z < 0$) that creates the main phase of magnetic storm. Every cross-section of the 3D-spirally contracting and expanding magnetosphere has a vortical field pattern that is likely to be followed by the space plasma to create, depending on its size, the dynamics of the plasmasphere, the vortices in the magnetopause boundary and the observed swirls of the auroral forms.

**Conclusion**

The found 3D-spiral expansions and contractions of the magnetosphere [15] are further developed with the revealed role of the solar wind dynamic pressure in the creation of the global depression of the geomagnetic field associated with magnetic storm. Other controversial issues of solar wind-magnetosphere interaction are also elucidated. The obtained results are summarised below:

a) The solar wind and plasma sheet particles are repelled at inner (outer) magnetic lines in the expanded (contracted) magnetosphere and so aurora is created at lower (higher) latitudes.
b) The greater dynamic pressure of the solar wind ($\rho V^2$) is balanced at inner magnetic lines and so a thicker magnetotail is created as the magnetosphere expands downstream. The northern lobe of the thicker magnetotail accumulates more effectively the southward IMF component ($B_s$) and becomes even thicker as the magnetosphere expands (Figs. 1, 2). So the negative $D_{st}$ excursion deepens and creates the main phase of magnetic storm. In this way the puzzling success of coupling functions of the type $\rho V^2 B_s$ in the $D_{st}$ prediction and the existence of threshold strength for $B_s$ that initiates or intensifies magnetic storm are explained.
c) The larger (smaller) magnetic expansion creates the main phase of magnetic storm (the growth phase of substorm). This accounts for the poorly understood similarity between intense magnetic substorms and weak storms. The expanded magnetotail contracts to create the expansion phase of magnetic substorm and to explain the puzzling near Earth location of the expansion onset.
d) The faster magnetic contraction creates a wider boundary layer between the outer low latitude magnetosphere and the ambient space plasma (Eq. 2). Particles from this layer are driven inward and create the intensification and poleward expansion of the auroral arcs, e.g. during the expansion phase of magnetic substorm.
e) The calculated average width ($0.44R_E$) and the modelled behaviour of the dayside low latitude magnetopause boundary layer (Eq. 2) are in agreement with observations. The low latitude magnetopause boundary layer becomes thinner (wider) in the expanding (contracting) magnetosphere.
f) Prolonged steady northward IMF strongly contracts the magnetosphere. Then pre-noon and post-noon dents will appear on the magnetopause because magnetic flux from these local time sectors will be carried by the solar wind downstream to produce the four lobes of the bifurcated magnetotail.
g) The magnetosphere expands and contracts and so accumulates solar wind particles and IMF flux. The acquired energy is delivered to smaller scales, seen as substorm expansions and intensification of the auroral activity.

The magnetosphere expands and contracts as a whole and in parts. In this manner it self-similarly and self-consistently evolves under the qualitatively different impacts of the solar wind dynamic pressure and the IMF (Figs. 1-3). The described simple magnetospheric behaviour is in accordance with the basic principle of parsimony, known also as Occam's razor [e.g., 34]. This fundamental principle of science requires not doing with more, e.g. with more assumptions, what can be done with less. Therefore it gives priority to the simpler model, i.e. supports the found simple magnetospheric behaviour.